\documentclass[showpacs,preprintnumbers,amsmath,amssymb,twocolumn,superscriptaddress,prb]{revtex4}
\usepackage{times}
\usepackage{graphicx}
\usepackage{amsfonts}
\usepackage{amsmath, amsthm, amssymb}
\usepackage{dsfont}
\usepackage{subfigure}

\newcommand{\vp}{\mathbf{p}} 
\newcommand{\vpf}{\mathbf{p}_\text{F}}
\newcommand{\vecr}{\mathbf{R}}
 
\newcommand{\e}[1]{\mathrm{e}^{#1}}
\newcommand{\g}{\check{g}(\vpf,\vecr;\varepsilon,t)}
\newcommand{\G}{\check{G}(\vp,\vecr;\varepsilon,t)} 
\newcommand{\ug}{\underline{g}}
\newcommand{\uf}{\underline{f}}

\newcommand{\eg}{\textit{e.g. }}
\newcommand{\etal}{\emph{et al. }}
\def\i{\mathrm{i}}

\begin{document}
\title[Spin-flip scattering and non-ideal interfaces in dirty ferromagnet/superconductor junctions]{Spin-flip scattering and non-ideal interfaces in dirty ferromagnet/superconductor junctions}
\author{Jacob Linder}
\affiliation{Department of Physics, Norwegian University of
Science and Technology, N-7491 Trondheim, Norway}
\author{Asle Sudb{\o}}
\affiliation{Department of Physics, Norwegian University of
Science and Technology, N-7491 Trondheim, Norway}

\date{Received \today}
\begin{abstract}
We study the proximity-induced superconducting correlations as well as the local density of states of a 
ferromagnet, in a ferromagnet/$s$-wave superconductor heterostructure. We include the effects of spin-flip 
scattering, non-ideal interfaces, and the presence of impurities in the sample. We employ the quasiclassical 
theory of superconductivity, solving the Usadel equation with emphasis on obtaining transparent analytical 
results. As our main result, we report that in a certain parameter regime the spatial oscillations of the 
anomalous (superconducting) part of the Green's function induced in the ferromagnet by the proximity effect 
from the $s$-wave superconductor, are damped out due to the presence of spin-flip processes. As a consequence, 
spin-flip scattering may under certain conditions actually enhance the local density of states due to the 
oscillatory behaviour of the latter in ferromagnet/superconductor structures. We also conjecture that the 
damping could be manifested in the behaviour of the critical temperature ($T_c$) of the $s$-wave superconductor 
in contact with the ferromagnet. More specifically, we argue that the non-monotonic decrease of $T_c$ in ferromagnet/$s$-wave superconductor junctions without magnetic impurities is altered to a monotonic, 
non-oscillatory decrease when the condition $1>16\tau_\text{sf}^2h^2$ is fulfilled, where $\tau_\text{sf}$ is 
the spin-flip relaxation time and $h$ is the exchange field.
\end{abstract}

\maketitle

\section{Introduction}
Proximity structures consisting of ferromagnetic and superconducting materials offer a synthesis between two important physical phenomena that may hold the potential for future applications in nanotechnology: spin-polarization and dissipationless flow of a current. Ferromagnetism is antagonistic to conventional superconductors, since the exchange field acts as a depairing agent for spin-singlet Cooper pairs. However, the proximity effect does not merely suppress the spin-singlet superconducting order parameter, but may also induce long-ranged spin-triplet correlations under certain circumstances \cite{bergeretRMP}. Much effort has been invested over the last decade to unveil various  physical phenomena that occur in ferromagnet/superconductor (F/S) heterostructures \cite{buzdin}. Among the highlights of such phenomena, it is natural to mention the $\pi$-state that is realized in S/F/S structures, which has been studied intensively both theoretically \cite{pijunctionst1,pijunctionst2,pijunctionst3} and experimentally \cite{ryazanovPRL01,pijunctionse}. In this state, the superconducting order parameters differ in sign in contrast to the usual $0$-state in S/N/S structure. The transition from a $0$- to $\pi$-state may be controlled by the width of the ferromagnet separating the superconductors, thus offering a way of manipulating the Josephson supercurrent that occurs in such systems. Another way of obtaining a $\pi$-state makes use of misaligned exchange fields in S/F heterostructures. This opportunity arises in a variety of systems, ranging from superconductors with spiral magnetic order \cite{kulic,eremin,champel}, thin S/F bilayers \cite{bergeretJOS,golubovJETP, li2002,linder3}, and so-called ferromagnetic superconductors \cite{gronsleth,linderprb2007} where ferromagnetic and superconducting order seem to coexist uniformly. The latter is most often interpreted as evidence for triplet pairing in the superconducting sector.
\par
Although various theoretical idealizations allow for a relatively simple approach to F/S heterostructures in the quasiclassical framework, the presence of factors such as non-ideal interfaces and both magnetic and non-magnetic impurities should be taken into account to obtain more precise agreement between theory and experiment. A particularly interesting feature in such hybrid structures is the generation of a spin-triplet superconducting component in the ferromagnet which survives even in the dirty limit due to a special symmetry property which was first suggested by Berezinskii \etal~\cite{bulaevskii}, and later predicted to occur in F/S junctions by \cite{bergeretPRL1,bergeretPRL2} Bergeret \etal This issue has been the subject of intense investigations during the past decade (see for instance Refs.~\onlinecite{bergeretRMP,buzdin} and references therein). In particular, the role of triplet pairing in superconductor/half-metal/superconductor structures has received much attention lately \cite{eschrig1,eschrig2,kopu,asano}, much due to the experimental verification of a Josephson current in 
such a setup \cite{klapwijk}. 
\par
\begin{figure}[h!]
\centering
\resizebox{0.46\textwidth}{!}{
\includegraphics{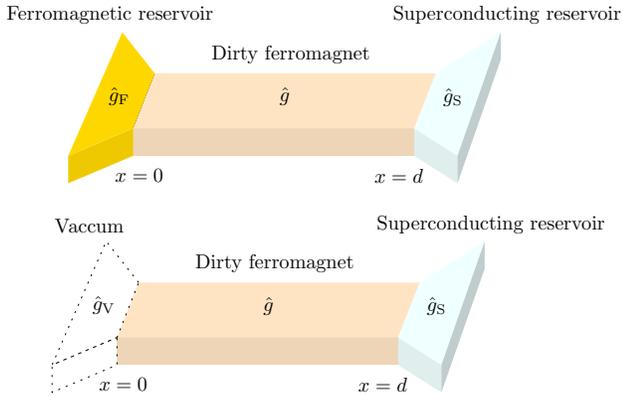}}
\caption{(Color online) The figure shows geometries that will be considered in this paper. In the top 
figure, we consider a F/F/S junction consisting of a dirty ferromagnet sandwiched between a ferromagnetic 
and superconducting reservoir where the Green's functions are described by their bulk values. In the bottom 
figure, we consider a F/S junction consisting of a dirty ferromagnet connected to a superconducting reservoir.}
\label{fig:setup}
\end{figure}

With regard to F/S junctions, two recent publications have adressed some aspects of how spin-flip processes 
affect the critical temperature \cite{gosu1} and the density of states \cite{gosu2}. Here, we will consider 
two different geometries to study the impact of spin-flip scattering and non-ideal interfaces in heterostructures involving ferromagnets and superconductors. The geometry of the systems we study are given in Fig. \ref{fig:setup}
In the top figure, we consider a dirty ferromagnet of width $d$ sandwiched between a ferromagnetic and 
superconducting reservoir, where the Green's functions are assumed to be in their bulk form. In the bottom 
figure, the ferromagnetic reservoir is replaced with vacuum, effectively leading to a F/S junction. 
\par
In this paper, we study the influence of magnetic impurities and non-ideal interfaces on the spatial- 
and energy-dependence of the anomalous (superconducting) part of the quasiclassical Green's functions 
induced in the ferromagnet by the proximity effect from the $s$-wave superconductor. In particular, we 
investigate how this is manifested in the local density of states (LDOS). We present analytical results 
that may elucidate features obtained numerically in Ref.~\onlinecite{gosu2}. In agreement with 
Ref.~\onlinecite{gosu1}, we find that spin-flip processes alters the decay and oscillation length of the 
anomalous Green's function in the ferromagnet. Our main result is that under certain conditions, the usual 
oscillations of the Green's function in F/S junctions without magnetic impurities vanish completely when 
the condition $1>16\tau_\text{sf}^2h^2$ is fulfilled, where $\tau_\text{sf}$ is the spin-flip relaxation 
time and $h$ is the exchange field. Defining $\Gamma = 1/\tau_\text{sf}$, an equivalent statement is 
to say that the oscillations vanish when the energy associated with spin-flip scattering exceeds a critical 
value $\Gamma_c = 4h$. As a direct consequence, the spin-flip scattering may actually enhance the LDOS due 
to the oscillations of the anomalous Green's function in the ferromagnet.  A detailed study is performed 
concerning how the length scales associated with the decay and the oscillations of the Green's function 
are affected by magnetic impurities. This is important in the context of understanding the behaviour of 
for instance the Josephson current in S/F/S structures, since spin-flip scattering will always be present 
to some degree in real samples. Including such effects will presumably yield a more satisfactory 
quantitative agreement with experimental data.
\par
We organize this paper as follows. In Sec. \ref{sec:theory}, we establish the theoretical framework we will use to treat the F/S hybrid structure. Namely, we employ the Keldysh formalism in the quasiclassical approximation to study 
the Usadel equation with appropriate boundary conditions at the interfaces. In Sec. \ref{sec:results}, we present 
our results for the spatial- and energy-dependence of the anomalous Green's function in the dirty ferromagnet, as 
well as results for the local density of states, both without (for reference) and with spin-flip scattering. We 
also discuss how the decay length and oscillation length scales of the Green's function are affected by the 
spin-flip scattering, providing transparent analytical results. In Sec. \ref{sec:discussion} and \ref{sec:summary}, 
we discuss and summarize the main results of the paper. We will use boldface notation for 3-vectors, $\check{\ldots}$ for $8\times8$ matrices, $\hat{\ldots}$ for $4\times4$ 
matrices, and $\underline{\ldots}$ for $2\times2$ matrices.

\section{Theoretical formulation}\label{sec:theory}
\subsection{Quasiclassical theory}
The central quantity in the quasiclassical theory of superconductivity  is the quasiclassical Green's functions 
$\g$, which depends on the momentum at Fermi level $\vpf$, the spatial coordinate $\vecr$, energy measured 
from the chemical potential $\varepsilon$, and time $t$. A considerable literature covers the Keldysh formalism 
and non-equilibrium Green's functions \cite{serene, kopnin, rammer, zagoskin}. Here we only briefly sketch the theoretical structure, for the sake of readability and for establishing notation. The quasiclassical Green's 
functions $\g$ is obtained from the Gor'kov Green's functions $\G$ by integrating out the dependence on kinetic 
energy, assuming that $\check{G}$ is strongly peaked at Fermi level,
\begin{equation}\label{eq:quasiclassical}
\g = \frac{\i}{\pi} \int \text{d}\xi_\vp \G.
\end{equation}
This above is typically applicable to superconducting systems where the characteristic length scale 
of the perturbations present, such as mean-free path and magnetic coherence length, is much smaller than the 
Fermi wavelength. Also, the corresponding characteristic energies of such phenomena must be much smaller than 
the Fermi energy $\varepsilon_\text{F}$. The quasiclassical Green's functions may be divided into an advanced 
(A), retarded (R), and Keldysh (K) component, each of which has a $4\times4$ matrix structure in the combined particle-hole and spin space.  One has that
\begin{equation}
\check{g} = \begin{pmatrix}
\hat{g}^\text{R} & \hat{g}^\text{K}\\
0 & \hat{g}^\text{A} \\
\end{pmatrix},
\end{equation}
where the elements of $\g$ read
\begin{equation}
\hat{g}^\text{R,A} = \begin{pmatrix}
\ug^\text{R,A} & \uf^\text{R,A}\\
-\tilde{\uf}^\text{R,A} & -\tilde{\ug}^\text{R,A} \\
\end{pmatrix},\;
\hat{g}^\text{K} = \begin{pmatrix}
\ug^\text{K} & \uf^\text{K}\\
\tilde{\uf}^\text{K} & \tilde{\ug}^\text{K} \\
\end{pmatrix}.
\end{equation}
The quantities $\ug$ and $\uf$ are $2\times2$ spin matrices, with the structure
\begin{equation}
\ug = \begin{pmatrix}
g_{\uparrow\uparrow} & g_{\uparrow\downarrow} \\
g_{\downarrow\uparrow} & g_{\downarrow\downarrow} \\
\end{pmatrix}.
\end{equation}
Due to internal symmetry relations between these Green's functions, all of these quantities are not independent. In particular, the tilde-operation is defined as
\begin{equation}
\tilde{f}(\vpf,\vecr;\varepsilon,t) = f(-\vpf,\vecr;-\varepsilon,t)^*.
\end{equation}
The quasiclassical Green's functions $\g$ may be determined by solving the Eilenberger \cite{eilenberger} equation
\begin{equation}\label{eq:eilenberger}
[\varepsilon\hat{\rho}_3 - \hat{\Sigma}, \check{g}]_\otimes + \i\mathbf{v}_\text{F}\boldsymbol{\nabla} \check{g} = 0,
\end{equation}
where $\hat{\Sigma}$ contains the self-energies in the system such as impurity scattering, superconducting order parameter, and exchange fields. The star-product $\otimes$ is noncommutative and is defined in Appendix \ref{star-product}. When there is no explicit time-dependence in the problem, the star-product reduces to normal multiplication. This is the case we will consider throughout the paper. The operation $\hat{\rho}_3\check{g}$ inside the commutator should be understood $
\hat{\rho}_3\check{g} \equiv \text{diag}\{ \hat{\rho}_3,\hat{\rho}_3\} \check{g}$.
Pauli-matrices in particle-hole$\times$spin (Nambu) space are denoted as $\hat{\rho}_i$, while Pauli-matrices in spin-space are written as $\underline{\tau}_i$. The Green's functions also satisfy the normalization condition 
\begin{equation}
\check{g}\otimes\check{g} = \check{1}.
\end{equation}
The self-energies entering Eq. (\ref{eq:eilenberger}) must be solved in a self-consistent manner. For instance, a weak-coupling $s$-wave superconducting order parameter is obtained by
\begin{equation}
\Delta(\vecr;t) = - \frac{\lambda}{4} \int^{\omega_\text{c}}_{-\omega_\text{c}} \text{d}\varepsilon \langle f_{\uparrow\downarrow}^\text{K} (\vpf, \vecr;\varepsilon,t)\rangle_{\hat{\mathbf{p}}_\text{F}},
\end{equation}
where $\omega_\text{c}$ is the cut-off energy, which may be eliminated in favor of the transition temperature. The notation $\langle \ldots \rangle$ is to be understood as an angular averaging over the Fermi surface. 
Once $\g$ has been determined, physical quantities of interest may be calculated, such as the electrical current
\begin{equation}\label{eq:current}
\mathbf{j}(\vecr;t) = \frac{N_\text{F}ev_\text{F}}{4} \int \text{d}\varepsilon \text{Tr}\{ \langle \hat{\rho}_3 \mathbf{e}_\text{F} \hat{g}^\text{K} \rangle_{\hat{\mathbf{p}}_\text{F}} \},
\end{equation}
where $N_\text{F}$ is the density of states (DOS) per spin at Fermi level. Eq. (\ref{eq:current}) also includes the contribution to charge transport for holes, thus including processes such as Andreev reflection. In the special case of an equilibrium situation, one may express the Keldysh component in terms of the retarded and advanced Green's function by means of the relation
\begin{equation}\label{eq:kra}
\hat{g}^\text{K} = (\hat{g}^\text{R} - \hat{g}^\text{A})\tanh(\beta\varepsilon/2),
\end{equation}
where $\beta = T^{-1}$ is inverse temperature. In nonequilibrium situations, one must derive kinetic equations for nonequilbrium distribution functions in order to specify the Keldysh part \cite{belzigreview}.
\par
The above equations suffice to completely describe for instance a single superconducting structure, but must be supplemented with boundary conditions when treating heterostructures such as F/S junctions. These boundary conditions take different forms depending on the physical properties of the interface, and we proceed to describe possible scenarios in this respect.
Transport across interfaces in heterostructures may in general be characterized according to three particular properties: \textit{i)} the transmission of the interface, \textit{ii)} the resistivity of the compounds separated by the interface, and \textit{iii)} whether the interface is spin-active or not. Let us clarify the distinction between the two first properties. The transmission of the barrier (assuming for simplicity a single open transport-channel) determines whether one is dealing with a point-contact or tunneling contact, which differ in terms of the likelihood of electron transport to occur across the interface. In the Blonder-Tinkham-Klapwijk-language \cite{btk}, the point-contact corresponds to low values of $Z$ while the tunneling limit is obtained for high values of $Z$. On the other hand, the resistivity of the compounds separated by the interface is unrelated to the transmissivity of the interface, and one may have for instance a tunneling contact with electrodes attached to it that have either a large or small resistance.
\par
The third property determines to what degree the interface discriminates between incoming quasiparticles with different spins. Zaitsev \cite{zaitsev} derived boundary conditions for a clean N/S interface, while Kuprianov and Lukichev (KL) \cite{kupluk} worked out simplified boundary conditions in the dirty limit, valid for atomically sharp interfaces in the tunneling regime with a low barrier transparency. Although the KL boundary conditions are strictly speaking not valid for high transparency of the barrier, they may be used for qualitative predictions in that regime under certain conditions \cite{lambert}. The most compact way of writing the boundary conditions for the Green's functions for arbitrary interfaces was introduced by Nazarov \cite{nazarov}. In all the preceding references, a non-magnetic (spin-inactive) interface was assumed. The generalized boundary conditions for magnetically active interfaces have also been derived \cite{millis}. Let us make a final remark concerning the treatment of interfaces in the quasiclassical theory of superconductivity. We previously stated that the present theory is valid as long as characteristic energies of various self-energies and perturbations in the system are much smaller than the Fermi energy. At first glance, this might seem to be inreconcilable with the presence of interfaces, which represent strong perturbations varying on atomic length scales, clearly in stark contradiction to the regime of validity of quasiclassical theory. However, this problem may be overcome by including the interfaces as boundary conditions for the Green's functions rather than directly in the Eilenberger equation.
\par
The KL boundary conditions may be applied for a dirty junction in the tunneling limit when the transparency of the interface is low, in correspondance with our assumption of a weak proximity effect. For the retarded part of the Green's function, they read
\begin{align}\label{eq:boundary}
2d\gamma(\hat{g}\partial_x\hat{g})\Big|_{x=0} = -[\hat{g},\hat{g}_{F(V)}]\Big|_{x=0},\notag\\
2d\gamma(\hat{g}\partial_x\hat{g})\Big|_{x=d} = [\hat{g},\hat{g}_{S}]\Big|_{x=d},
\end{align}
where F (V) corresponds to the F/F/S (F/S) case of Fig. \ref{fig:setup}. The parameter $\gamma$ models the interfacial transmission properties, and is given by $\gamma = R_\text{I}/R_\text{F}$ where $R_\text{I}$ is the interface resistance per unit area, while $R_\text{F}$ is the equivalent in the dirty ferromagnet. In this work, $\gamma$ holds the status of a variational parameter. A low transparency of the interface amounts to the regime $\gamma\gg 1$. As previously mentioned, qualitative predictions may still be obtained using the linearized Usadel equations for $\gamma \simeq 1$, and even the quantative aspects of the exact numerical solution may in some cases be very well modelled by this approximation \cite{hammer}. Under the assumption of a weak proximity effect, we will neglect the depletion of the superconducting order parameter near the interface in order to facilitate the calculations and for the sake of obtaining analytical results. Moreover, we will use the bulk solution of the Green's function in the superconductor. This approximation is valid when the superconducting region is much less disordered than the ferromagnet \cite{bergeretRMP}.

\subsection{Green's functions}
We will consider the dirty limit of the Eilenberger equation Eq. (\ref{eq:eilenberger}), which leads to the Usadel equation \cite{usadel}. This will be an appropriate starting point for diffusive systems where the scattering time due to impurities satisfies $X\tau \ll 1$, where $X$ is the energy scale of any other self-energy in the problem. For strong ferromagnets where $h$ becomes comparable to $\varepsilon_\text{F}$, the stated inequality may strictly speaking not be valid for $X=h$. Hence, we  will restrict ourselves to the regime $h\ll \varepsilon_\text{F}$. Below, we will mostly concern ourselves with the retarded part of $\g$, since the advanced component may be found via the relation
\begin{equation}\label{eq:gagr}
\hat{g}^\text{A} = -(\hat{\rho}_3 \hat{g}^\text{R} \hat{\rho}_3)^\dag.
\end{equation}
The Keldysh component is calculated by means of Eq. (\ref{eq:kra}) in a situation of local thermal equilibrium. For a non-equilbrium situation, the Keldysh component is found by 
\begin{equation}\label{eq:keldyshnoneq}
\hat{g}^\text{K} = \hat{g}^\text{R}\hat{\mathcal{F}} - \hat{\mathcal{F}}\hat{g}^\text{A},
\end{equation} where $\hat{\mathcal{F}}$ is a matrix distribution function to be determined from kinetic equations, while Eq. (\ref{eq:gagr}) still holds. Eq. (\ref{eq:keldyshnoneq}) follows from the normalization condition of the Green's function, and the distribution function may be chosen as diagonal without any loss of generality. In general, we may write
\begin{equation}
\hat{\mathcal{F}} = \mathcal{F}_l\hat{1} +\mathcal{F}_t\hat{\rho}_3,
\end{equation}
where comparison with Eq. (\ref{eq:kra}) shows that in a thermal equilibrium one has $\mathcal{F}_l = \tanh(\beta\varepsilon/2)$, $\mathcal{F}_t = 0$.
\par
By isotropizing the Green's function due to the assumed frequent impurity scattering, it is rendered independent of $\vpf$. This isotropic (in momentum space) Green's function satisfies the Usadel equation in the ferromagnet:
\begin{equation}\label{eq:usadel}
D\nabla (\check{g}\nabla\check{g}) + \i[\varepsilon\hat{\rho}_3 + \hat{M} - \check{\sigma}_\text{sf}, \check{g}] = 0.
\end{equation}
Above, the exchange energy $h$ is accounted for by the matrix $\hat{M} = \text{diag}(h\underline{\tau_3},h\underline{\tau_3})$, assuming a magnetization in the $\mathbf{z}$-direction, while the spin-flip self-energy reads
\begin{equation}
\check{\sigma}(\vecr;\varepsilon) = -\frac{\i}{8\tau_\text{sf}} \sum_i \hat{\alpha}_i \check{g}(\vecr;\varepsilon)\hat{\alpha}_i,
\end{equation}
where $\tau_\text{sf}$ is the spin-flip scattering time. We have defined the matrices $\hat{\alpha}_i = \text{diag}(
\underline{\tau_i},\underline{\tau_i}^\text{T})$. The diffusion constant is given by $D=v_\text{F}^2\tau/3$.
Although the Usadel equation in general requires a numerical solution, an analytical approach is permissable under certain conditions. In the case of a weak proximity effect, one may effectively linearize Eq. (\ref{eq:usadel}). This is a valid treatment for low transparency interfaces or close to $T_c$. In this case, Eq. (\ref{eq:usadel}) is expanded around the bulk solution. To be definite, let us consider the retarded part of Eq. (\ref{eq:usadel}) which has the same form, namely
\begin{equation}\label{eq:usadelR}
D\nabla (\hat{g}^\text{R}\nabla\hat{g}^\text{R}) + \i[\varepsilon\hat{\tau}_3 + \hat{M} - \hat{\sigma}_\text{sf}, \hat{g}^\text{R}] = 0,
\end{equation}
where $\hat{\sigma}_\text{sf}$ is obtained from $\check{\sigma}_\text{sf}$ simply by letting $\check{g} \to \hat{g}^\text{R}$. Omitting the superscript on the Green's function, we may expand it around the bulk solution $\hat{g}_0$ as $\hat{g} \simeq \hat{g}_0 + \hat{f}$, where $\hat{g}_0 = \text{diag}(\bar{1},-\bar{1})$ and
\begin{align}
\hat{f} &= \begin{pmatrix} 
\bar{0} & \bar{f}(\vecr;\varepsilon) \\
-[\bar{f}(\vecr;-\varepsilon)]^* & \bar{0}\\
\end{pmatrix},\notag\\
 \bar{f}(\vecr;\varepsilon) &= 
\begin{pmatrix}
f_{\uparrow\uparrow}(\vecr;\varepsilon) & f^\text{t}_{\uparrow\downarrow}(\vecr;\varepsilon)+(^{\text{t}\to\text{s}})\\
f_{\uparrow\downarrow}^\text{t}(\vecr;\varepsilon)-(^{\text{t}\to\text{s}})& f_{\downarrow\downarrow}(\vecr;\varepsilon) \\
\end{pmatrix}
\end{align}
One may now multiply out the matrix equation Eq. (\ref{eq:usadelR}), only keeping the lowest order terms in the anomalous Green's functions $f_{\alpha\beta}(\vecr;\varepsilon)$. For more compact notation, we define the quantities
\begin{align}
f^\text{i}_{\alpha\beta}\equiv f^\text{i}_{\alpha\beta}(\vecr;\varepsilon),\; \{\alpha,\beta\} = \uparrow,\downarrow \text{ and }\; f^\text{t(s)}\equiv f_{\uparrow\downarrow}^\text{t(s)},\; \text{i}=\text{s,t}.\notag\\
\end{align}
We will proceed to consider the two distinct cases illustrated in Fig. \ref{fig:setup}. The Green's functions in the different reservoirs read 
\begin{align}
\hat{g}_\text{V} = \hat{0},\;
\hat{g}_\text{F} = \begin{pmatrix}
\bar{1}& \bar{0}\\
\bar{0}&\bar{1}\\
\end{pmatrix},\; \hat{g}_\text{S} =  \begin{pmatrix}
\bar{1}c & \i\bar{\tau}_2s\\
\i\bar{\tau}_2s & - \bar{1}c\\
\end{pmatrix},
\end{align}
where we have defined $c\equiv\text{cosh}(\vartheta)$, $s\equiv\text{sinh}(\vartheta)$ with $\vartheta \equiv \text{atanh}(|\Delta|/\varepsilon)$. Note that we have set the superconducting phase equal to zero, thus considering a gauge where the gap is a purely real quantity.

\subsection{Odd-frequency pairing}

Before moving on to the graphical presentation of our results, let us comment on the presence of the $S_z=0$ triplet component of the anomalous Green's function in the ferromagnet. It is well-known that even in the absence of spin-flip processes $(\tau_\text{sf}\to\infty)$, the triplet component is generated in the ferromagnet due to the presence of the exchange field $h$. We will later investigate how the magnitude of this triplet component is affected by including spin-flip processes. Also, it is of interest to investigate the symmetry properties of the singlet and triplet component. Since we are considering the isotropic part (with respect to momentum) of the Green's function due to the angular averaging in the dirty limit, one would naively expect that only the singlet component should be present. This is because the singlet anomalous Green's function is usually taken to be even under inversion of momentum, while the triplet components are taken to be odd under inversion of momentum. Recall that inversion of momentum amounts to an exchange of spatial coordinates for the field operators, since $\mathbf{p}$ is the Fourier transform of the relative coordinate $\mathbf{r}\equiv \mathbf{r}_1 - \mathbf{r}_2$. However, another possibility exists that permits for the presence of triplet correlations in the ferromagnet, namely a sign shift under inversion of energy. This type of pairing has been dubbed \textit{odd-frequency pairing} in the literature, interpreting energy as a real frequency. Recall that inversion of energy is equivalent to an exchange of time coordinates for the field operators, since $\varepsilon$ is the Fourier transform of the relative time coordinate $t\equiv t_1-t_2$. For a detailed discussion of even- and odd-frequency pairing, the reader may consult Appendix \ref{even-odd}.
\par
Let us in passing show that the singlet component is even in frequency, while the triplet component is odd in frequency, by using the definition in Eq. (\ref{eq:defevenodd}). To do this, we must first find the advanced Green's function $f^\text{A}$ by exploiting Eq. (\ref{eq:gagr}). Direct matrix multiplication leads to
\begin{align}
\hat{g}^\text{A} &=
\begin{pmatrix}
-1 & 0 & 0 & -f_-^\text{R}(-\varepsilon) \\
0 & -1 & -f_+^\text{R}(-\varepsilon) \\
0 & [f_-^\text{R}(\varepsilon)]^* & 1 & 0 \\
[f_+^\text{R}(\varepsilon)]^* & 0 & 0 & 1 \\
\end{pmatrix}.
\end{align}
From this, one infers that $f_\text{s}^\text{A}(\varepsilon) = f_\text{s}^\text{R}(-\varepsilon)$ (even-frequency pairing) and $f_\text{t}^\text{A}(\varepsilon) = -f_\text{t}^\text{R}(-\varepsilon)$ (odd-frequency pairing). We have included this short paragraph on even- and odd-frequency pairing to emphasize that although even-frequency triplet correlations are destroyed in the dirty limit due to the isotropization stemming from impurity scattering, odd-frequency triplet correlations may persist since these do not vanish under angular averaging. Also, it is important to understand that the triplet pairing we are discussing in the present paper is then quite different from the triplet pairing in for instance Sr$_2$RuO$_4$. In the latter case, the triplet pairing is odd in momentum and therefore even in frequency \cite{maeno}. As a consequence, superconductivity is highly sensitive to impurity scattering in Sr$_2$RuO$_4$ and only observed in very clean samples.

\par
In the problem under consideration, the two important energies are the exchange energy $h$ and the BCS gap $|\Delta|$. Associated with these energies are two typical length scales: the correlation length in the ferromagnet $\xi_\text{F} = \sqrt{D/h}$ and the superconducting coherence length $\xi_\text{S} = \sqrt{D/(2\pi T_\text{c})}$, where the critical temperature in a weak-coupling superconductor is given by $|\Delta| \simeq 1.76 T_\text{c}$. One may think of $\xi_\text{F}$ is the penetration depth of the superconducting condensate into the dirty ferromagnet. In an experimental situation, one usually has $h\gg|\Delta|$ even for relatively weak ferromagnets, such that $\xi_\text{F} \ll \xi_\text{S}$. For the quasiclassical treatment to be valid, one must then have $|\Delta| \ll h \ll \varepsilon_\text{F}$. For a Fermi energy of 1 eV, it would then be reasonable to consider $h$ in the neighborhood of 30 meV and $|\Delta|$ around 1 meV. The effect of $D$ and $d$ may be accounted for in the single parameter $\varepsilon_\text{T} = D/d^2$, named the Thouless energy. This is the relevant energy scale for the proximity effect in the case of highly transparent interfaces. In the following, we will unless specifically stated otherwise fix $h/|\Delta| = 30$ to operate within the allowed boundaries of our approximations. Since one is often interested in investigating how various physical properties behave as a function of the thickness $d$ of the ferromagnetic layer, it is useful to note that for $d/\xi_\text{S} = x$, one finds
\begin{equation}
\varepsilon_\text{T} = \frac{2\pi|\Delta|}{1.76x^2}.
\end{equation}

\section{Results}\label{sec:results}
We now provide the main results of this paper, namely a study of how the triplet correlations and LDOS are 
affected by spin-flip scattering in a F/S and F/F/S junction. When including scattering upon magnetic 
impurities in the sample, Eq. (\ref{eq:usadel}) yields the differential equations 
\begin{align}\label{eq:diffspinflip}
&D\partial_x^2(f_\text{t} \pm f_\text{s}) + 2\i(\varepsilon\pm h) (f_\text{t} \pm f_\text{s}) - \frac{1}{2\tau_\text{sf}}(f_\text{t} \pm 3f_\text{s}) = 0,\notag\\
&D\partial_x^2 f_\sigma + (2\i \varepsilon - \frac{1}{2\tau_\text{sf}}) f_\sigma = 0.
\end{align}
Note that we have here assumed an isotropic spin-flip disorder, in contrast to the uniaxial disorder 
considered in Refs.~\onlinecite{faure2,oboznov2, gosu1}. We comment more on this in Sec. \ref{sec:discussion}.
Spin-flip processes in combination with a spatially homogeneous exchange field do not lead to equal-spin 
correlations in the ferromagnet, although the inclusion of a spin-active barrier will generate these 
components \cite{kopu, linder307}. Therefore, for the present case of a non-magnetic interface, we 
have that $f_\sigma=0$. For the $S_z=0$ triplet and singlet Green's functions, the general solution of 
Eq. (\ref{eq:diffspinflip}) reads
\begin{align}
f_\text{t} &= c_1 \e{-q_+x} + c_2\e{q_-x} + c_3\e{q_+x} + c_4\e{-q_-x},\notag\\
f_\text{s} &= \frac{\i}{4\tau_\text{sf}h}(c_1\kappa_- \e{-q_+x} + c_2\kappa_+\e{q_-x} \notag\\
&\hspace{0.5in}+ c_3\kappa_-\e{q_+x} + c_4\kappa_+\e{-q_-x}),
\end{align}
where we have defined 
\begin{align}
q_\pm &= \Big[-\Big(4\i \tau_\text{sf} \varepsilon - 2 \pm \sqrt{1 - 16\tau_\text{sf}^2h^2}\Big)/(2D\tau_\text{sf})\Big]^{1/2},\notag\\
\kappa_\pm &= 1 \pm \sqrt{1 - 16\tau_\text{sf}^2h^2}.
\end{align}
The coefficients $\{c_i\}$ will be determined from the boundary conditions of the F/S and F/F/S junctions. These are given by Eq. (\ref{eq:boundary}), which may be written in terms of the $f_\pm$ functions. For the F/S junction, we have
\begin{align}\label{eq:boundsf1}
\textit{i): }& \partial_x f_\pm\Big|_{x=0} = 0,\notag\\
\textit{ii): }& d\gamma \partial_x f_\pm \Big|_{x=d} = \pm s - cf_\pm\Big|_{x=d}.
\end{align}
In the F/F/S case, the condition \textit{ii)} is still valid while \textit{i)} must be replaced with
\begin{align}\label{eq:boundsf2}
\textit{i): } d\gamma \partial_x f_\pm \Big|_{x=0} = f_\pm\Big|_{x=0}.
\end{align}
Note that it is implicit here that $f_\pm = f^\text{R}_\pm$. The resulting analytical expressions or $\{c_i\}$ read as follows:
\begin{align}
c_4 &= \frac{-2s(1+X_-/X_+)}{Y_- - Y_+X_-/X_+},\; c_3 = (2s - c_4Y_+)/X_+, \notag\\
c_2 &= Rc_4,\; c_1 = L_2c_2 + L_3c_3 + L_4c_4.
\end{align}
For convenience, we have defined the following quantities:
\begin{align}
\mathcal{A}_\pm &= 1\pm \i\kappa_-/(4\tau_\text{sf}h),\; \mathcal{B}_\pm = 1\pm \i\kappa_+/(4\tau_\text{sf}h),\notag\\
L_2 &= \frac{\mathcal{B}_+(\psi+q_-)}{\mathcal{A}_+(q_+-\psi)},\; L_3 = \frac{\psi+q_+}{q_+-\psi},\notag\\
L_4 &= \frac{\mathcal{B}_+(\psi-q_-)}{\mathcal{A}_+(q_+-\psi)},\; R = \frac{q_- - \psi}{q_-+\psi},
\end{align}
in addition to 
\begin{align}
X_\pm &= \mathcal{A}_\pm[\e{q_+d}(2c+2\gamma dq_+) + \e{-q_+d}L_3(2c-2\gamma dq_+)],\notag\\
Y_\pm &= \mathcal{B}_\pm[\e{q_-d}R(2c+2\gamma dq_-) + \e{-q_-d}(2c-2\gamma dq_-)]\notag\\
&+ \mathcal{A}_\pm\e{-q_+d}(L_4 + L_2R)(2c-2\gamma dq_+).
\end{align}
In the F/S case, $\psi=0$, while $\psi=(-\gamma d)^{-1}$ in the F/F/S case.
The knowledge of these coefficients completely determines the spatial- and energy-dependence of the anomalous Green's functions everywhere in the dirty ferromagnet. When applying the limit $\tau_\text{sf}\to\infty$, by making use of 
\begin{equation}
\mathcal{A}_\pm \to 2\delta_{\pm,+},\; \mathcal{B}_\pm \to 2\delta_{\pm,-},\; q_\pm \to ik_\pm,
\end{equation}
one regains well-known results for the scenario without spin-flip scattering.
Also, it is worth noting that the triplet component vanishes for $h=0$, even in the presence of spin-flip scattering. Having calculated the Green's functions, we may now study the effect of spin-flip scattering on the LDOS.
The spin-resolved LDOS is given by 
\begin{align}\label{eq:DOS}
N_\sigma(\vecr;\varepsilon) &= N_{\text{F},\sigma} \text{Re}\Bigg\{ \Big(1 + [f_\text{t}(\varepsilon) + \sigma f_\text{s}(\varepsilon)]\notag\\
&\times[f_\text{t}(-\varepsilon) - \sigma f_\text{s}(-\varepsilon)]^* \Big)^{1/2}\Bigg\},\; \sigma=\uparrow,\downarrow=\pm1.
\end{align}
Above, $N_{\text{F},\sigma}$ is the DOS at Fermi level for spin species $\sigma$ ($N_\text{F} = \sum_\sigma N_{\text{F},\sigma}$). The deviation $\delta N$ from the bulk DOS inside the dirty ferromagnet may be defined as
\begin{equation}
\delta N \equiv \sum_\sigma [N_\sigma(\vecr;\varepsilon) - N_{\text{F},\sigma}]/N_{\text{F},\sigma}.
\end{equation}
We also define the normalized LDOS as $N = \sum_\sigma N_\sigma(\vecr;\varepsilon)/(2N_{\text{F},\sigma})$, such that in the absence of a proximity effect, $N=1$.
The oscillations of the LDOS in a F/S junction was first reported by Buzdin \cite{buzdinbrief}, and lead to observable effects such as the nonmonotonic dependence of the critical temperature on the length $d$ of a F/S bilayer \cite{lazar} and the $\pi$-phase structures that occur in F/S hybrid systems \cite{ryazanovPRL01}.
\par
\begin{figure}[h!]
\centering
\resizebox{0.5\textwidth}{!}{
\includegraphics{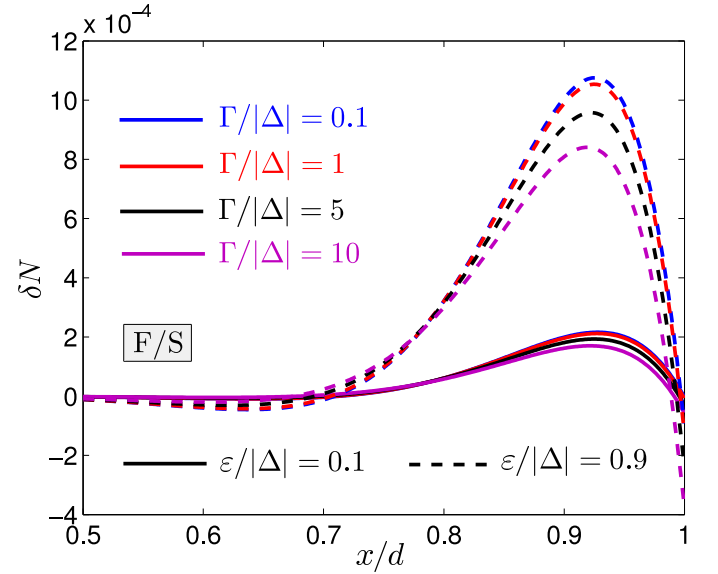}}
\caption{(Color online) Spatial variation of the deviation from the LDOS ($\delta N$) for a F/S junction in the presence of spin-flip scattering. We have chosen $\varepsilon_\text{T}/|\Delta| = 1$ and $\gamma=5$. The qualitative features are the same for the F/F/S junctions for this particular set of parameters. It is seen that for a given energy, increasing spin-flip scattering $\Gamma = 1/\tau_\text{sf}$ will increase the oscillation length and reduce the amplitude of the Green's function.}
\label{fig:DOSxSF}
\end{figure}

Consider first Fig. \ref{fig:DOSxSF} for a plot of the correction $\delta N$ to the LDOS as a function of position in the dirty ferromagnet. Although the maximum amplitude of $\delta N$ is suppressed with increasing spin-flip scattering $\Gamma = 1/\tau_\text{sf}$, an interesting feature is that the magnitude of the correction ($|\delta N|$) is in some regions actually enhanced due to spin-flip scattering. This seems to be a result of the oscillatory behaviour of the LDOS. In a N/S junction, where there is no oscillatory behaviour of the LDOS, spin-flip scattering would simply cause a reduction of the correction $\delta N$. Thus, the role of spin-flip scattering in a F/S junction is more subtle than in a N/S junction where it simply amounts to a suppression of the LDOS.
\par 
It might seem counter-intuitive that increasing spin-flip scattering should increase the correction to the LDOS, since the anomalous Green's functions should be suppressed for large $\Gamma$. We suggest the following resolvement of this phenomena. It is clear that the LDOS displays an oscillatory behaviour due to the presence of an exchange field, both with and without the spin-flip scattering. However, in the presence of spin-flip processes, the period of these oscillations is modified. From Fig. \ref{fig:DOSxSF}, it is seen that the peak of the correction to $\delta N$ is suppressed with increasing $\Gamma$. But even though this peak becomes smaller, the different periods of oscillation allows $|\delta N(\Gamma_1)|$ to outgrow $|\delta N(\Gamma_2)|$ at certain distances from the interface, even for $\Gamma_1 > \Gamma_2$. This is a subtle feature unique for F/S interfaces in the presence of spin-flip processes as compared to N/S junctions.
\par
It is interesting to investigate the role of spin-flip scattering with regard to the decay and oscillating lengths further. Very recently, some aspects of this topic were adressed in Ref.~\onlinecite{gosu3}. We here examine in detail some features that occur when spin-flip scattering is included in a F/S junction, among them the \textit{vanishing} of the characteristic oscillations of the anomalous Green's function $\hat{f}$ in a certain parameter regime. Consider first the case without spin-flip scattering, effectively letting $\tau_\text{sf}\to\infty$. From our previous equations, it is clear that if we write
\begin{equation}
k_\pm = \sqrt{2\i(\varepsilon\pm h)/D} = k_{1,\pm} + \i k_{2,\pm},
\end{equation}
then the real quantities $k_{1,\pm}$ and $k_{2,\pm}$ correspond to the oscillating part and decaying part of $f_\pm$, respectively. We ignore the $\varepsilon$-term since we consider the regime $h\gg \varepsilon$, and write $|k_{1,\pm}| = 1/\xi_\text{osc}$, $|k_{2,\pm}| = 1/\xi_\text{dec}$. One readily obtains
\begin{align}
\xi_\text{dec} = \xi_\text{osc} = \xi_\text{F}.
\end{align}
In other words, we recover the well-known fact that the oscillating and decaying length scales of the superconducting condensate in the absence of spin-flip scattering are equal \cite{buzdin}. Consider now a finite value of $\tau_\text{sf}$, where we obtain
\begin{align}\label{eq:qeq}
q_\pm = \sqrt{\frac{2 \mp \sqrt{1 - 16\tau_\text{sf}^2h^2}}{2D\tau_\text{sf}}}.
\end{align}
We have neglected the energy term, assuming $\varepsilon\tau_\text{sf} \ll 1$. In this case, writing $q_\pm = q_{1,\pm} + \i q_{2,\pm}$ means that $q_{1,\pm}$ and $q_{2,\pm}$ are associated with the decay and oscillating length, respectively. We may now distinguish between two cases. If the inequality
\begin{equation}\label{eq:ineq}
1>16\tau_\text{sf}^2h^2
\end{equation}
is satisfied, then $q_\pm$ is purely real. In this case, there are \textit{no oscillations} of the anomalous Green's function in the ferromagnet. Since we assumed that $\varepsilon\tau_\text{sf} \ll 1$, this means that the exchange field should be sufficiently weak for the vanishing of oscillations to take place. For instance, given a spin-flip energy of $\Gamma/|\Delta| = 20$ one would need $h/|\Delta| < 5$ for the oscillations to disappear. If the spin-flip energy becomes very large, then the oscillations would vanish even for moderate exchange fields. This prediction should have easily observable experimental consequences, manifested for instance in the behaviour of the critical temperature as a function of junction width $d$, given that the required parameter regime may be experimentally realized. We comment further on this later in this paper.
\par
\begin{figure}
\centering
\subfigure[\text{ } Plot of the characteristic decay and oscillation lengths $(\xi_x)$ of the anomalous Green's function in the presence of spin-flip scattering with $h/|\Delta| = 5$.] 
{
    \label{fig:decay}
    \includegraphics[width=0.5\textwidth]{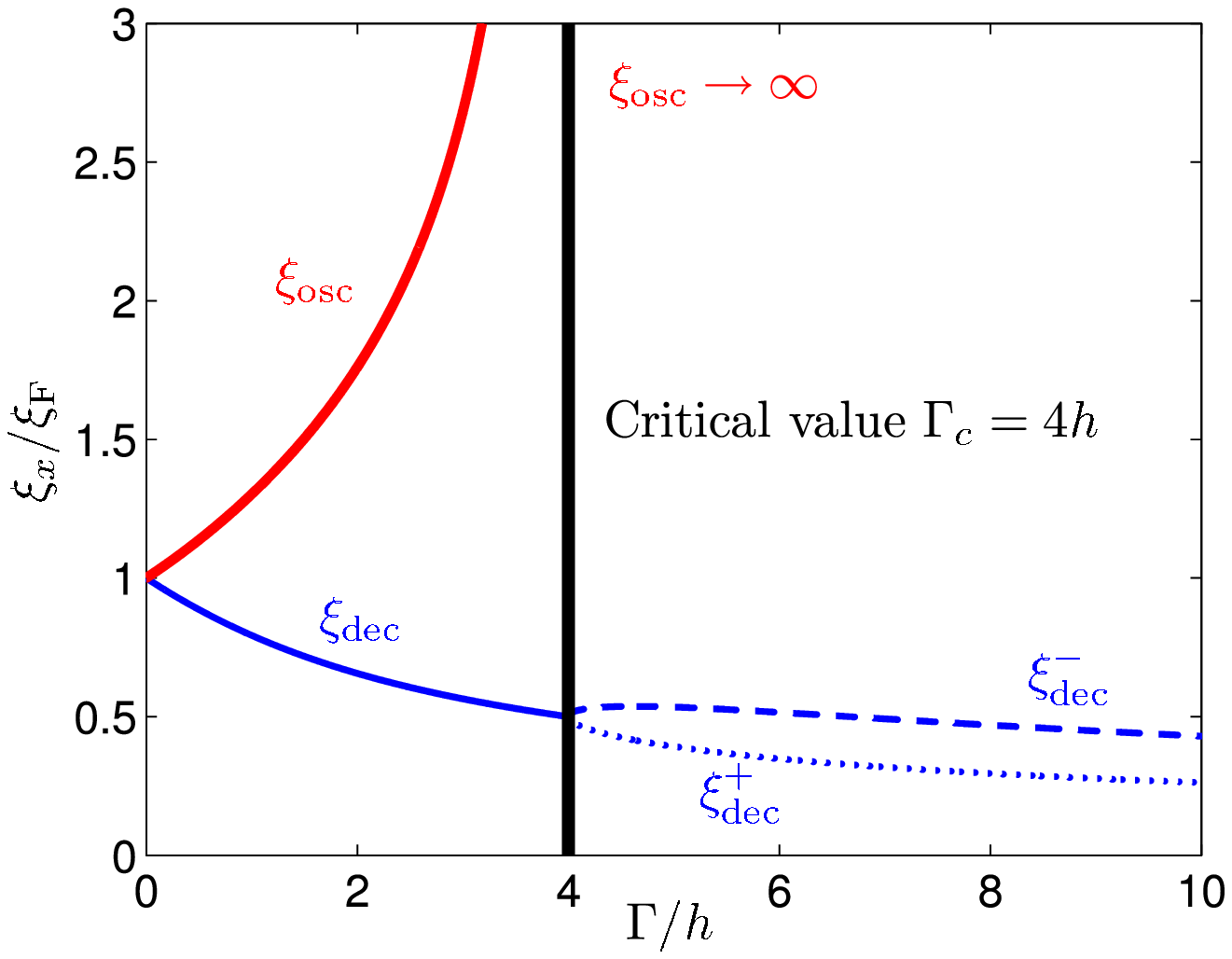}
}
\hspace{1cm}
\subfigure[\text{ } Spatial variation of the deviation from the LDOS ($\delta N$) for a F/S junction in the presence of spin-flip scattering. We have chosen $\varepsilon_\text{T}/|\Delta| = 1$, $\gamma=5$, and $h/|\Delta|=10$. The quasiparticle energy has been set to $\varepsilon/|\Delta|=0.5$, but the qualitative behaviour is identical for all $\varepsilon<|\Delta|$. ] 
{
    \label{fig:decay2}
    \includegraphics[width=0.5\textwidth]{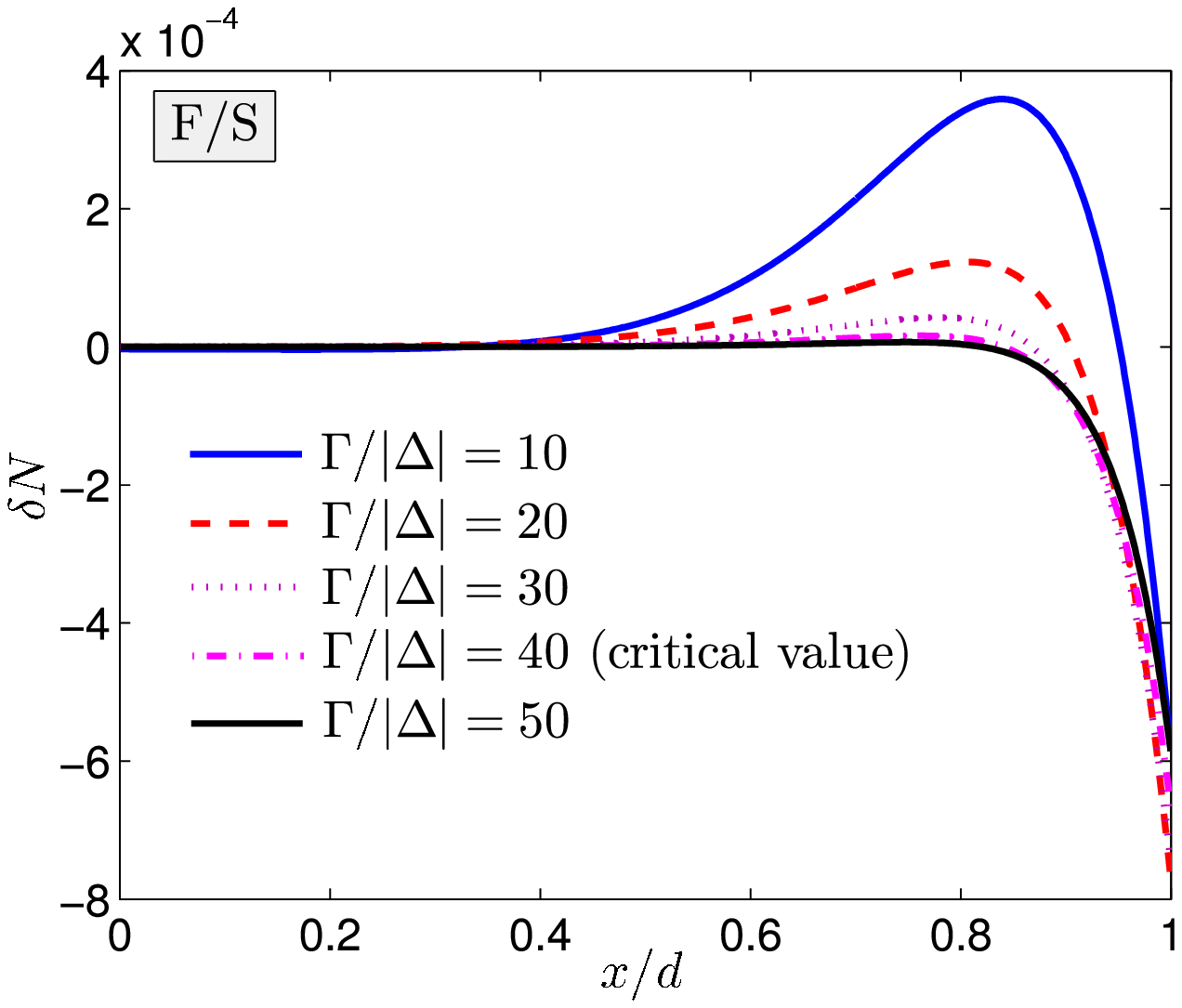}
}
\caption{(Color online) As shown in (b), there are no oscillations of the Green's function in the ferromagnet in the parameter range $\Gamma/h>4$ ($\Gamma = 1/\tau_\text{sf}$). Inclusion of the energy contribution $\varepsilon$ brings small corrections to this result, but as seen from the figure the condition Eq. (\ref{eq:ineq}) is a very good approximation. This behaviour is to be contrasted with the usual oscillations in F/S junctions as obtained without spin-flip scattering. Note that in (a), $\xi_\text{osc}$ formally diverges near $\Gamma=\Gamma_c$ which separates the two parameter regimes where oscillations occur and where they do not occur.}
\label{fig:decay} 
\end{figure}

In the case where $1<16\tau_\text{sf}^2h^2$, $q_\pm$ is no longer purely real, and oscillations are again present in $f_\pm$. It is instructive to consider a plot of $\xi_\text{osc}$ and $\xi_\text{dec}$ as a function of spin-flip scattering to see how the oscillation and decay length are affected by these processes. This is done in Fig. \ref{fig:decay}, where the divergence of the oscillation length is clearly seen at the critical value $\Gamma_c= 4h$. Note that as $\Gamma\to 0$, $\xi_\text{osc}$ and $\xi_\text{dec}$ become equal, as previously stated. When Eq. (\ref{eq:ineq}) is satisfied, the decay length is different for the up- and down-spins. This may be seen by introducing $\xi^\pm_\text{dec} = 1/q_\pm$ as defined by Eq. (\ref{eq:qeq}). Our results for $\Gamma < \Gamma_c$ are consistent with Ref.~\onlinecite{gosu1}, who reported that increased spin-flip scattering should increase the oscillation length and reduce the decay length. Let us consider how this effect is manifested in the LDOS, a directly measurable experimental quantity. In Fig. \ref{fig:decay2}, we plot the spatial correction to the LDOS for $h/|\Delta|=5$ for several values of the spin-flip energy. As seen, the oscillations vanish as $\Gamma$ increases, and are completely absent when Eq. (\ref{eq:ineq}) is satisfied. The effect we predict should thus be measurable via scanning-tunneling microscopy (STM) measurements, by probing the LDOS.

\par
\begin{figure}[h!]
\centering
\resizebox{0.5\textwidth}{!}{
\includegraphics{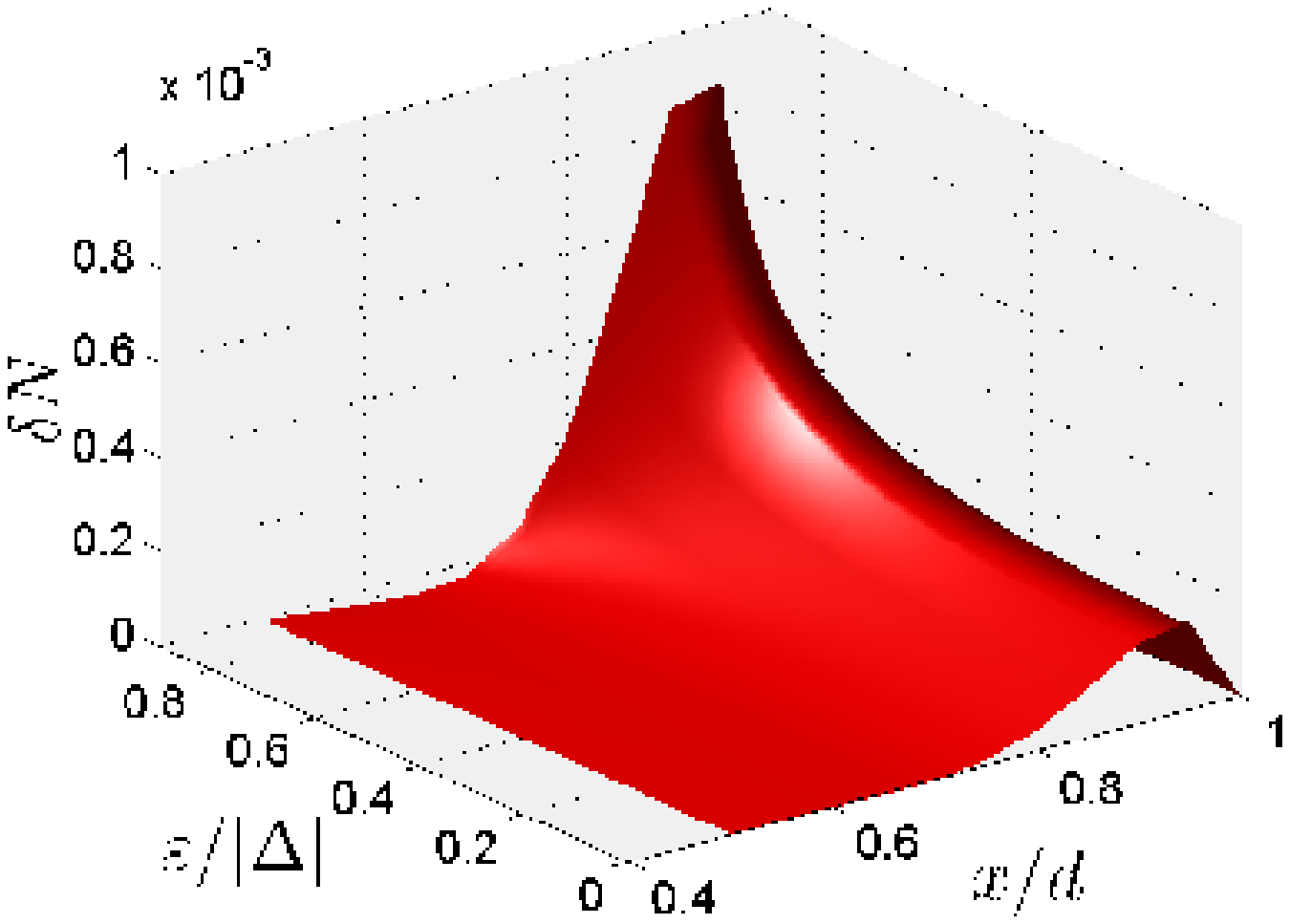}}
\caption{(Color online)  Correction to the LDOS ($\delta N$) for $\Gamma/h = 0.01$ ($\Gamma = 1/\tau_\text{sf}$). Surface plot of the deviation from the LDOS in the $(x,\varepsilon)$-plane for a junction of width $d=2\xi_\text{S}$ and with transparency parameter $\gamma=5$ and exchange field $h/|\Delta|=30$. The most protruding feature is the peak emerging in the LDOS right at the F/S interface, followed by a dip-structure at low energies. }
\label{fig:DOScontGamma1}
\end{figure}

We next consider a surface plot of the correction to the LDOS in the $(x,\varepsilon)$-plane for a set of parameters that should correspond to a quite typical F/S junction. We set $h/|\Delta|=30$ and $\Gamma/|\Delta|=0.3$. We use the analytical results for the F/S case, since the difference from a F/F/S junction is small in the low transparency regime.
As seen in Fig. \ref{fig:DOScontGamma1}, the LDOS peaks at $x=d$ with an amplitude that increases with energy. This peak vanishes upon increasing the spin-flip scattering parameter $\Gamma$, corresponding to the effect we predict - namely that the oscillations seen in the LDOS of a F/S junction vanish above a critical value of the spin-flip scattering energy.

\section{Discussion}\label{sec:discussion}
In F/S junctions without magnetic impurities, it is well-known that the critical temperature of the superconductor exhibits a non-monotonic dependence on the thickness of the ferromagnet layer $d$ (for an extensive treatment of this topic, see the review by Buzdin \cite{buzdin}) . The physical reason for the damped, oscillatory behaviour of $T_c$ in such systems is not completely understood. It seems reasonable to attribute this characteristic feature to the oscillatory behaviour of the Green's functions in the ferromagnet, since this non-monotonic behaviour of $T_c$ is not observed in N/S junctions \cite{jiang, ogrin}. However, the transparency of the barrier may also play a key role in the manifestion of oscillations in $T_c$, as argued in for instance Ref.~\onlinecite{aarts}. In another experiment, a purely monotonic decay of $T_c$ as a function of $d$ was observed in a Pb/Ni junction \cite{bourgeoisanddynes}. Even though the physical picture is not fully resolved, it is highly plausible that the oscillations of $\hat{f}$ are intimately linked to the behaviour of $T_c$. In this paper, we have shown that the presence of spin-flip scattering may significantly change the qualitative behaviour of $\hat{f}$ in the ferromagnet. Specifically, the oscillations vanish at a critical value $\Gamma_c=4h$. Under the assumption that the characteristic behaviour of $\hat{f}$ strongly influences the fashion in which $T_c$ decays, it is then clear that one should observe a transition from a non-monotonic (damped, oscillating) to a pure monotonic (damped) decay of $T_c$ upon increasing the concentration of magnetic impurities in a sample. Our finding offers a new, possible explanation for the experiments where a monotonic decay of $T_c$ was observed, namely that the concentration of magnetic impurities was such that the critical value $\Gamma_c$ was exceeded.
\par
In this paper, we have assumed an isotropic spin-flip disorder, in contrast to the uniaxial magnetic disorder considered in Refs.~\onlinecite{faure2,oboznov2,gosu1}. This leads to somewhat different equations for the proximity-induced anomalous Green's function in the ferromagnet. Since we have considered a strongly uniaxial 
exchange field, it is implicitly assumed that the presence of spin-flip scattering in the plane perpendicular 
to the exchange field still allows for the uniaxial field to be accomodated. In the case of strong
uniaxial anisotropy, the first of the Usadel equations Eq. (\ref{eq:diffspinflip}) is replaced
by \cite{oboznov2}
\begin{equation}
\label{diffequniaxial}
D\partial_x^2(f_\text{t} \pm f_\text{s}) + 2\i(\varepsilon\pm h) (f_\text{t} \pm f_\text{s}) - \frac{1}{2\tau_\text{sf}}(f_\text{t} \pm f_\text{s}) = 0.
\end{equation}
Note how the factor $3$ in the last term in Eq. (\ref{eq:diffspinflip}) now is replaced by unity.
Following the same line of reasoning that led to Eq. (\ref{eq:qeq}), we now obtain
\begin{equation}
q_\pm = \sqrt{\frac{1 \mp 4 \i \tau_{\rm{sf}} h}{2 D \tau_\text{sf}}}.
\end{equation}
This quantity is always complex, and hence we  always obtain (damped) oscillations and never a 
complete supression of the oscillations. Thus, the model with isotropic scattering and strongly 
uniaxially anisotropic scattering are qualitatively different. The model with isotropic scattering 
Eq. (\ref{eq:diffspinflip}) is expected to be most relevant for a weak exchange field, while the 
model Eq. (\ref{diffequniaxial}) is expected to be most relevant for  strong uniaxial anisotropy 
\cite{oboznov2}. Clearly, it would also be interesting to investigate a model which interpolates 
between these two limits, in order to investigate at what maximum anisotropy in the scattering a 
complete suppression of oscillations can occur. One may consider this situation crudely by  
the following Usadel equation
\begin{equation}
\label{diffequniaxial}
D\partial_x^2(f_\text{t} \pm f_\text{s}) + 2\i(\varepsilon\pm h) (f_\text{t} \pm f_\text{s}) - \frac{1}{2\tau_\text{sf}}(f_\text{t} \pm \beta f_\text{s}) = 0
\end{equation}
where we have introduced the parameter $\beta$ to account for the unixial ($\beta=1$) and the isotropic 
($\beta =3$) case. Again, following the line of reasoning that led to Eq. (\ref{eq:qeq}), we now find 
\begin{align}\label{eq:qbeta}
q_\pm = \sqrt{\frac{(1+\beta)/2 \mp \sqrt{(\beta - 1)^2/4 - 16\tau_\text{sf}^2h^2}}{2D\tau_\text{sf}}}.
\end{align}
From this simple analysis, one would tentatively conclude that the case of strong uniaxial
anisotropy is special, in that it is the only case where one cannot possibly  obtain 
$(\beta - 1)^2/4 - 16\tau_\text{sf}^2h^2 =0$ for any finite $\tau_\text{sf}$ and $h$, which 
is the requirement for suppression of oscillations. For all other  values of $\beta$, it would 
be possible to obtain a real $q$ and hence complete supression of oscillations. Clearly, this 
matter warrants further and detailed investigations. 

\par
In recent publications \cite{crouzy1,crouzy2}, Crouzy \etal considered the interesting problem of a Josephson current in a S/F/F'/S 
structure with non-collinear magnetizations in the ferromagnetic regions. It was shown that the misorientation angle between the ferromagnetic layers could be used to progressively shift the junction between a $0$- and $\pi$-state. In deriving their results, effects such as spin-flip scattering and non-ideal interfaces were omitted for simplicity. Our analytical results  account for both of these effects, and may thus be useful to obtain a quantitatively better agreement for the Josephson effect 
with experimental data by including these phenomena in S/F/F'/S structures. Work in this direction is now in progress \cite{future}.

\section{Summary}\label{sec:summary}
In conclusion, we have investigated the role of spin-flip scattering and non-ideal interfaces in dirty ferromagnet/superconductor (F/S) junctions. Our analytical results may serve as a basis for calculating other physical quantities of interest in F/S multilayers, such as the Josephson current. The main result of this paper is that  we show analytically how the well-known oscillations of the anomalous Green's function induced in the ferromagnet by the superconductor in F/S structures without magnetic impurities vanish completely above a critical value for the energy associated with spin-flip scattering, $\Gamma$. More precisely, we find that the oscillations are absent when the condition $\Gamma_c>4h$ is fulfilled, where $h$ is the exchange field. We have argued that this might be experimentally observable through a transition from a non-monotonic (damped, oscillating) to a monotonic (damped) decrease of the critical temperature of the junction as a function of the thickness of the ferromagnet layer.
As another consequence, we find that increasing spin-flip scattering may actually enhance the local density of states (LDOS) under certain conditions. This is a quite subtle effect that might seem counter-intuitive at first glance. However, it may be understood by realizing that the anomalous Green's functions display an oscillatory behaviour in the presence of an exchange field. The period of these oscillations is modified in the presence of spin-flip scattering. This means that although the amplitude of the oscillations decreases for increasing spin-flip scattering, the correction to the LDOS may in certain spatial intervals actually be greater for large spin-flip scattering than for weak spin-flip scattering. 
This finding may be of importance in order to correctly interpret LDOS-spectra as obtained from \eg scanning tunneling microscopy measurements.

\acknowledgments
One of the authors (J. L.) is greatly indebted to J. P. Morten for providing him with a detailed introduction to the quasiclassical theory of superconductivity. D. Huertas-Hernando and H. J. Skadsem are also thanked for useful comments. This work was supported by the Norwegian Research Council Grant Nos. 158518/431, 158547/431, (NANOMAT), 
and 167498/V30 (STORFORSK).

\appendix

\section{\label{star-product} Defining the star-product}
\noindent We here define the star-product which enters the Eilenberger equation Eq. (\ref{eq:eilenberger}). For any two functions $A$ and $B$, we have
\begin{equation}
A\otimes B = \e{\i(\partial_{T_A}\partial_{\varepsilon_B} - \partial_{\varepsilon_A}\partial_{T_B})/2}AB,
\end{equation}
where the differentiation operators denote derivation with respect to the variables $T$ and $\varepsilon$ in the mixed representation. Note that if there is no explicit time-dependence in the problem, the star-product reduces to regular multiplication.

\section{\label{even-odd} Even- and odd-frequency pairing}
\noindent Consider the anomalous Green's function with time-ordering operator $\mathcal{T}$,
\begin{equation}\label{eq:start}
f_{\alpha\beta}(\mathbf{r}_1,\mathbf{r}_2;t_1,t_2) = -\i\mathcal{T}\{ \langle \psi_\alpha(\mathbf{r}_1;t_1) \psi_\beta(\mathbf{r}_2;t_2)\rangle\},
\end{equation}
which in the mixed representation may be written as
\begin{equation}
f_{\alpha\beta}(\mathbf{r}_1,\mathbf{r}_2;t_1,t_2) = f_{\alpha\beta}(\mathbf{R},\mathbf{r};T,t).
\end{equation} 
The Pauli-principle introduces restrictions on this correlation function for equal times $t_1=t_2=t'$, namely
\begin{equation}\label{eq:paulione}
f_{\alpha\beta}(\mathbf{r}_1,\mathbf{r}_2;t') = -f_{\beta\alpha}(\mathbf{r}_2,\mathbf{r}_1;t').
\end{equation}
This follows directly from the anticommutation relation for the field operators in Eq. (\ref{eq:start}), and reflects the fact that the fermionic two-particle anomalous Green's function must be antisymmetric under exchange of particle coordinates. Assume now for ease of notation that there is no explicit time-dependence in the problem and that the system is homogeneous, which allows us to discard the dependence on the center-of-mass coordinates $\mathbf{R}$ and $T$. The following argumentation is valid even if this simplification may not be performed, and the equations then hold for each set of points $(\mathbf{R},T)$. By a Fourier-transform, we now obtain 
\begin{align}
f_{\alpha\beta}(\vp;t) &= \int \text{d}\mathbf{r}\e{-\i\vp\mathbf{r}} f_{\alpha\beta}(\mathbf{r};t).
\end{align}
The Pauli-limitation Eq. (\ref{eq:paulione}) then reads $f_{\alpha\beta}(\vp;0) = f_{\beta\alpha}(-\vp;0)$, since equal times give $t=0$. This seems to indicate that the Green's function must be odd under inversion of momentum or exchange of spin coordinates. However, another possibility exists, as may be seen by Fourier-transforming 
\begin{align}
f_{\alpha\beta}(\vp;\varepsilon ) &= \int \text{d}t \e{\i \varepsilon t} f_{\alpha\beta}(\vp;t).
\end{align}
In terms of the momentum- and energy-dependent Green's functions, the Pauli-principle now dictates that
\begin{align}
\int \text{d}\varepsilon  f_{\alpha\beta}(\vp;\varepsilon ) = -\int \text{d}\varepsilon  f_{\beta\alpha}(-\vp,\varepsilon ).
\end{align}
This gives us two possibilities that are still perfectly compatible with the equal-time restriction: either $f_{\alpha\beta}(\vp;\varepsilon ) = -f_{\beta\alpha}(-\vp;\varepsilon )$ or $f_{\alpha\beta}(\vp;\varepsilon ) = -f_{\beta\alpha}(-\vp;-\varepsilon )$. This is summarized in the equation
\begin{equation}\label{eq:paulicond}
f_{\alpha\beta}(\vp;\varepsilon ) = -f_{\beta\alpha}(-\vp;-\varepsilon ),
\end{equation}
which contains all possible symmetry classifications for the Green's functions that are compatible with the Pauli-principle. These are listed in Tab. \ref{tab:pauli}. Let us also make contact with the Matsubara formalism, 
where the anomalous Green's function is defined as
\begin{equation}
f^\text{M}_{\alpha\beta}(\mathbf{r}_1,\mathbf{r}_2;\tau_1,\tau_2) = -\mathcal{T}\{ \langle \psi_\alpha(\mathbf{r}_1;\tau_1) \psi_\beta(\mathbf{r}_2;\tau_2)\rangle\},
\end{equation}
and after a Fourier-transformation to the mixed representation satisfies
\begin{align}
f^\text{M}_{\alpha\beta}(\vp;\i\omega_n) &= \int^\beta_0 \text{d}\tau \e{\i\omega_n\tau} f^\text{M}_{\alpha\beta}(\vp;\tau),\notag\\
f^\text{M}_{\alpha\beta}(\vp;\tau) &= \frac{1}{\beta} \sum_n \e{-\i\omega_n\tau} f^\text{M}_{\alpha\beta}(\vp;\i\omega_n),
\end{align}
with $\tau$ as a complex time, $\beta$ as inverse temperature, and frequencies $\omega_n = (2n+1)\pi/\beta$. In this technique, one may apply the same procedure as for the real-time Green's functions and arrive at
\begin{align}
\sum_n [f^\text{M}_{\alpha\beta}(\vp;\i\omega_n) + f^\text{M}_{\beta\alpha}(-\vp;\i\omega_n)] = 0,
\end{align}
which also leads to the requirement that 
\begin{equation}\label{eq:matsubara}
f^\text{M}_{\alpha\beta}(\vp;\i\omega_n) = -f^\text{M}_{\beta\alpha}(-\vp;-\i\omega_n).
\end{equation}
The real-time retarded and advanced Green's functions may be obtained from the Matsubara Green's function by analytical continuation as follows $(\delta\to0)$:
\begin{align}
\lim_{\i\omega_n\to \varepsilon \pm\i\delta} f^\text{M}_{\alpha\beta}(\vp;\i\omega_n) = f^\text{R(A)}_{\alpha\beta}(\vp;\varepsilon).
\end{align}
\begin{table}[h!]
\centering{
\caption{Symmetry classifications of the anomalous Green's function that are compatible with the Pauli-principle. The "even" and "odd" terminology in the notation here points to the symmetry under a sign shift in energy, while "singlet" and "triplet" denotes the symmetry under exchange of spins. With these two properties in hand, the symmetry classification of momentum is given from the requirement that the entire function must be antisymmetric.}
	\label{tab:pauli}
	\vspace{0.15in}
	\begin{tabular}{ccccc}
	  	 \hline
	  	 \hline
		  \textbf{Spin} \hspace{0.1in} & \textbf{Momentum}	\hspace{0.1in} 	& \textbf{Energy} \hspace{0.1in} 	& \textbf{Notation}  \\
	  	 \hline
	  	Odd	\hspace{0.1in}	& Even	\hspace{0.1in}	& Even	\hspace{0.1in}	& Even singlet \\
	    Even	\hspace{0.1in}	& Odd	\hspace{0.1in}	& Even	\hspace{0.1in}	& Even triplet \\
	    Even	\hspace{0.1in}	& Even	\hspace{0.1in}	& Odd	\hspace{0.1in}	& Odd triplet \\
	    Odd	\hspace{0.1in}	& Odd	\hspace{0.1in}	& Odd	\hspace{0.1in}			& Odd singlet\\
			 \hline
	  	 \hline
	\end{tabular}}
\end{table}
From Eq. (\ref{eq:paulicond}), one infers that a triplet correlation may be even under momentum inversion \textit{if} it is odd under energy inversion. This scenario has been dubbed \textit{odd-frequency} pairing in the literature. The Pauli-principle can also be expressed by the retarded and advanced anomalous Green's functions by using Eq. (\ref{eq:matsubara}). To see this, we perform an analytical continuation on the right hand side of Eq. (\ref{eq:matsubara}), yielding
\begin{align}
\lim_{\i\omega_n \to \varepsilon+\i\delta} f^\text{M}_{\alpha\beta}(\vp;\i\omega_n) &= f^\text{M}_{\alpha\beta}(\vp;\varepsilon+\i\delta) \notag\\
&= f^\text{R}_{\alpha\beta}(\vp;\varepsilon),
\end{align}
while the same operation on the left-hand side produces
\begin{align}
\lim_{\i\omega_n \to \varepsilon+\i\delta} [-f^\text{M}_{\beta\alpha}(-\vp;-\i\omega_n)] &= -f^\text{M}_{\beta\alpha}(\vp;-\varepsilon-\i\delta) \notag\\
&= -f^\text{A}_{\beta\alpha}(-\vp;-\varepsilon).
\end{align}
Equating the two sides, we finally arrive at 
\begin{equation}\label{eq:pauli}
f^\text{R}_{\alpha\beta}(\vp;\varepsilon) = -f^\text{A}_{\beta\alpha}(-\vp;-\varepsilon).
\end{equation}
Actually, this information is embedded already in the definitions of the retarded and advanced Green's functions, and Eq. (\ref{eq:pauli}) may be verified by direct Fourier-transformation without going via Eq. (\ref{eq:matsubara}). It is also worth underscoring that the Matsubara technique is only valid for equilibrium situations, while the Keldysh formalism and the corresponding Green's functions is viable also for non-equilibrium situations. 
The distinction between odd- and even-frequency correlations for the retarded and advanced Green's functions is now as follows:
\begin{align}\label{eq:defevenodd}
\text{Odd-frequency:}&\; f^\text{R}_{\alpha\beta}(\vp;\varepsilon) = -f^\text{A}_{\alpha\beta}(\vp;-\varepsilon),\notag\\
\text{Even-frequency:}&\; f^\text{R}_{\alpha\beta}(\vp;\varepsilon) = f^\text{A}_{\alpha\beta}(\vp;-\varepsilon).
\end{align}

\end{document}